\documentclass[11pt,twoside]{article}

\usepackage{asp2014}

\aspSuppressVolSlug
\resetcounters

\begin{document}

\title{The International Virtual Observatory Alliance in 2018}

\author{Mark G. Allen,$^1$ Patrick Dowler,$^2$ Janet D. Evans,$^3$ Chenzhou Cui,$^4$ and Tim Jenness$^5$
for the IVOA Executive Committee and Technical Coordination Group}
\affil{$^1$Observatoire astronomique de Strasbourg, UMR 7550, F-67000 Strasbourg, France; \email{mark.allen@astro.unistra.fr}}
\affil{$^2$Canadian Astronomy Data Centre, National Research Council Canada, Victoria, British Columbia, Canada}
\affil{$^3$Center for Astrophysics, Harvard \& Smithsonian, Cambridge, MA, USA}
\affil{$^4$National Astronomical Observatories, CAS, Chaoyang District, 100101 Beijing, China}
\affil{$^5$Large Synoptic Survey Telescope, Tucson, AZ, USA}

\paperauthor{Mark G. Allen}{mark.allen@astro.unistra.fr}{0000-0003-2168-0087}{Observatoire astronomique de Strasbourg, UMR 7550,}{}{Strasbourg}{}{67000}{France}
\paperauthor{Patrick Dowler}{patrick.dowler@nrc-cnrc.gc.ca}{0000-0001-7011-4589}{Canadian Astronomy Data Centre}{National Research Council Canada}{Victoria}{British Columbia}{V9E 2E7}{Canada}
\paperauthor{Janet D. Evans}{janet@cfa.harvard.edu}{0000-0003-3509-0870}{Center for Astrophysics | Harvard & Smithsonian}{HEAD/CXC}{Cambridge}{MA}{02347}{USA}
\paperauthor{Chenzhou Cui}{ccz@bao.ac.cn}{0000-0002-7456-1826}{National Astronomical Observatories, CAS}{China-VO}{Chaoyang District}{Beijing}{100101}{China}
\paperauthor{Tim~Jenness}{tjenness@lsst.org}{0000-0001-5982-167X}{LSST}{Data Management}{Tucson}{AZ}{85719}{U.S.A.}



\begin{abstract}
The International Virtual Observatory Alliance (IVOA) held its bi-annual Interoperability Meeting over two and half days prior to the ADASS 2018 conference. We provide a brief report on the status of the IVOA and the activities of the Interoperability Meeting held in College Park.
\end{abstract}


\section{An alliance for the global vision of the Virtual Observatory}

The Virtual Observatory (VO) is a collection of interoperating data archives and software tools that facilitate astronomical research.  The overall goal is to support innovative research in astronomy by exploiting the full power of growing and emerging datasets and interoperable services. Many projects and data centres worldwide are working to make data and other resources work as a seamless whole.  The International Virtual Observatory Alliance\footnote{\url{http://www.ivoa.net}} (IVOA) is an organisation that debates and agrees on the technical standards that are needed to make the VO possible.  Constituted in 2002 \citep[see for example][]{2004SPIE.5493..137Q}, the IVOA has now been joined by 21 national and international VO projects that meet bi-annually. Major IVOA accomplishments include standards for data and metadata (Data Models), data exchange methods (Data Access Layer; Query Language), and a registry that lists available services and identifies what can be done with them.  Organisations have implemented VO-enabled tools and services that can interface seamlessly with VO-enabled archives worldwide, and some projects  have built new data systems with VO services at their core.  The IVOA acts as a focus for VO objectives, a framework for discussing and sharing VO ideas and technology, and a body for promoting and publicizing the VO.

Recent IVOA activities have focused on the engagement with future large data producing projects, and priorities have most recently focused on multi-dimensional data, and time domain astronomy. IVOA is also evolving as an organisation in a rapidly changing landscape, with the emergence of new large initiatives for research data sharing such as the Research Data Alliance \citep{doi:10.1087/20140503} with strong support of the \textbf{F}indable, \textbf{A}ccessible, \textbf{I}nteroperable, and \textbf{R}eusable \citep[FAIR;][]{doi:10.1038/sdata.2016.18} principles. Astronomy is also changing as we enter an era of very large data, and multi-wavelength and multi-messenger astrophysics where there is an essential need for high level interoperability of data, simulations, tools and services. The bi-annual Interoperability Meetings are working meetings for making progress on and facing the challenges \citep{2017ASPC..512...65A}  of coordinating the global effort for interoperability in Astronomy.

The IVOA work is pursued by Working Groups (WG) and Interest Groups (IG) (see table \ref{IVOA_wg_ig}), coordinated by the Technical Coordination Group (TCG), guided by a scientific priorities committee (CSP), with the overall direction provided by the IVOA Executive Committee. Participation in the IVOA working and interest groups is open, as is the participation in the bi-annual Interoperability meetings.

\section{IVOA November 2018}

The November 2018 Interoperability meeting\footnote{\url{https://wiki.ivoa.net/twiki/bin/view/IVOA/InterOpNov2018MeetingPage}} was held in College Park, MD, USA,  preceding ADASS XXVIII. One hundred and nine participants gathered for two and a half days of productive discussions.  Sessions were held by most of the WGs and IGs, running in two parallel streams, and with a number of joint WG-IG sessions.

\begin{table}[!ht]
\caption{IVOA Working Groups and Interest Groups \label{IVOA_wg_ig}}
\smallskip
\begin{center}
{\small
\begin{tabular}{@{}p{0.17\textwidth}p{0.8\textwidth}@{}}  
\tableline
\noalign{\smallskip}
Working Group & Description\\
\noalign{\smallskip}
\tableline
\noalign{\smallskip}
Applications &
Tools that Astronomers use to access VO data and services.
Standards specific to VO Astronomy-user-Applications. \\
\noalign{\smallskip}
Data Access Layer &
Define and formulate VO standards for remote data access.\\
\noalign{\smallskip}
Data Modelling &
Framework for the description of metadata attached to observed
or simulated data, and logical relationships between metadata. \\
\noalign{\smallskip}
Grid \& Web Services &
Grid technologies and web services within the VO context.\\
\noalign{\smallskip}
Semantics &
Explore technology in the area of semantics with the aim of
producing new standards that aid the interoperability of VO
systems. UCDs, Standard Vocabulary, exploration of Ontologies.\\
\noalign{\smallskip}
Resource Registry &
Registry provides the mechanism with which users and
applications discover and  select resources -- typically,
data and services.\\
\noalign{\smallskip}
\tableline 
\noalign{\smallskip}
Interest Group & Description\\
\noalign{\smallskip} 
\tableline
\noalign{\smallskip}
Time Domain &
Representation of the emerging time domain community,
specific time domain issues in a VO context.\\
\noalign{\smallskip}
Solar System &
IVOA standards in the scope of Solar System sciences. \\
\noalign{\smallskip}
Theory &
Ensuring that theoretical data and services are taken
into account in the IVOA standards.\\
\noalign{\smallskip}
Education &
VO tools, data and practices in support of astronomy teaching
in schools and universities. \\
\noalign{\smallskip}
Data Curation \& Preservation &
Share best practices and engage IVOA member projects
in the long-term curation and preservation of astronomical data.\\
\noalign{\smallskip}
Knowledge Discovery &
Knowledge discovery is the task of processing and analyzing
data-sets with the aim of extracting new knowledge. This area
spans visualization, remote data exploration, machine learning
techniques, statistical methods, workflow orchestration, and
polymorphic data access in the context of the VO.\\
\noalign{\smallskip}
Operations &
Coordinate and publicize activities of individuals, institutions and
groups interested in facilitating robust operations of distributed
astronomy applications, particularly those based upon
implementations of IVOA protocols.\\
\tableline\
\end{tabular}
}
\end{center}
\end{table}

 \subsection{Applications}
In the Applications (Apps) sessions the themes were: Python and its increasing importance in astronomy software and services; the VOTable format including its use in mapping of data models; HEALPix \citep{2005ApJ...622..759G} and its use in coverage maps (MOC) and the Hierarchical Progressive Survey \citep[HiPS;][]{2015A&A...578A.114F}; also the use of Authentication in applications and services. The relationship between Python and the VO was widely discussed and questions were raised about where best to contribute VO tools, how to raise the visibility of Python VO code, and how to avoid duplication. The discussions recognised a need for reference implementations of IVOA standards in Python, and potential follow-up activities such as IVOA hack-a-thons  were identified.
This is discussed further in \citet{O5-2_adassxxviii}.

\paragraph{Data Access Layer}
The Data Access Layer (DAL) WG sessions included discussions on the current minor version updates being prepared for the Table Access Protocol (TAP), Astronomical Data Query Language (ADQL) and DataLink standards that allow linking of datasets with various resources such as related datasets, metadata, or other services. New proposals were put forward for standards in support of multi-messenger astrophysics.  These standards are intended to describe the sky visibility of observatories and missions as is needed for the follow-up of events such as gravitational wave detections. The DAL sessions also included feedback on implementation of VO standards by NED and All-Sky Virtual Observatory \citep[ASVO;][]{O10-5_adassxxviii} groups.

\paragraph{Data Modelling}
The Data Modelling WG sessions reviewed the preparation of the major new version of the Space Time Coordinates (STC) standard. Progress was reported on the STC component models: coordinates, measurements, and transformations. A revision of the Simple Spectral Line standard was discussed. The concepts and purposes for the mapping of data models were reviewed. The Provenance data model standards document was in the "Request For Comments" (RFC) phase during the meeting, and presentations on the topic of provenance highlighted the reference implementations using the new model including one being used in the preparation of the Cherenkov Telescope Array (CTA).

\paragraph{Grid and Web Services}
The Grid and Web Services (GWS) sessions covered the topic of Authentication and Authorization, in particular the use of OAuth (Open Authorization) an open standard based on tokens.  The application of Big Data technologies was discussed, including an example of Apache Spark for simulated Euclid data. Feedback was provided on the use of the VOSpace 2.1 standard. The use of the VO Service Interface (VOSI) for managing versioning of services was discussed, and a new IVOA note is in preparation to outline best practices for its practical use.

\paragraph{Registry}
The Registry WG activities were largely combined into joint sessions with other WGs namely DAL, GWS and Apps. The role of the Registry in providing Authenticated Endpoints for resources was discussed, and a TAP prototype was shown. In the context of the VODataService standard, the use of spatial coverage maps (MOCs) in the registry was discussed. A plan was made to clean up the Registry-of-Registries (RofR)  to improve operations by removing invalid records.

\paragraph{Semantics}
The Semantics WG reviewed the organisation of the various Vocabularies, and discussed the needs expressed by the Theory IG, Data Model WG (for coordinate frame semantics), and Solar System IG. The nomenclature for describing instruments and facilities was also discussed.

\paragraph{Operations}
The Operations IG sessions included presentations on service monitoring and "weather reports" of VO services. VO service up-times are shown to be 98-99\% overall. VOParis and the European Space Agency (ESA) monitoring systems show that Cone Search services are now generally fully compliant with VO standards. Simple Image Access (SIA version 1) services have some common minor issues, as do the more complex Table Access Protocol (TAP) and and Simple Spectral Access (SSA) services. This level of monitoring is very valuable for the operation of the overall system, and also for highlighting areas of the standards that need improvement, indeed some issues may be most appropriately resolved by updating standards. Another topic was HTTPS, reporting that most VO protocols are compatible with HTTPS but challenges remain for WebSAMP (Web enabled Simple Application Messaging Protocol).

\paragraph{Time Domain Astronomy}
Time Domain astronomy is a current priority area for the IVOA. The Time Domain IG session addressed the status of the Time Series Data Model, and various ways of expressing time related data in the VO. Following the publication of an IVOA note, a proposal was presented for a TIMESYS element in VOTable, as a basic component for interoperability of time based data expressed in VOTable documents. The driving use case is the combination of two time-series, requiring two time coordinates to be put into a common frame of time scales  with a reference position and a time origin of the time scale. Consistency with future STC2 data model is taken into account.

\paragraph{Data Curation and Preservation}
The Data Curation and Preservation (DCP) session focused on the role of Digital Object Identifiers (DOIs), including examples of current practice and different approaches being taken throughout the astronomy and data sharing communities.
Vizier presented a proposal for how they were considering using DOIs for their holdings and how this might work for datasets where they were not the primary data source.
LSST discussed different ways in which DOIs for queries of large datasets could be issued and how to distinguish between the query itself, for a specific data release, and the results of the query, along with the possibility of allowing curated subsets of query results.
Finally it was agreed to look into issuing DOIs for VO standards.
A summary of the discussion was presented at ADASS \citep{B4_adassxxviii}.

\paragraph{Solar System}
 The Solar System IG sessions included reports from projects such as EuroPlanet VESPA, and there was discussion on bridging VESPA and PDS4. The use of the EPN-TAP protocol, a specialised version of TAP used in the planetary science context in the VESPA project, was discussed. A preliminary discussion was held about Space Reference Frames.

\paragraph{Theory}
The discussion in the Theory IG session highlighted the need to raise the visibility of Theory IVOA standards in upcoming projects that will make heavy use of simulations (EUCLID, LSST, SKA etc.), and to make stronger links between observations and theory.

\paragraph{ Knowledge Discovery and Education}
While the Knowledge Discovery IG  and Education IGs did not have sessions at this meeting, the members highlighted the various relevant current topics from across the various WGs and IGs. KDIG emphasised the need for Science Platforms for using and sharing scientific workflows and batch processing using the concept of "code to the data", and the use of DOIs. Education IG made plans for exploring connections to Education and Public Outreach activities of the IAU for VO-enabled data driven astronomical education.

\section{IVOA next steps}
The work of the IVOA is all aimed at enabling new and innovative science. The IVOA standards, defined \emph{by the community, for the community}, form an important part of the astronomy data infrastructure. Realising the vision of the VO of course relies on widespread implementation of the infrastructure via the adoption of the standards in data archives, services and tools. A recent visible success is the publication of the ESA Gaia mission DR2 via VO technologies, proving VO capabilities for handling the peak load challenges of a very large data release \citep[see e.g.,][]{ivoa_newsletter_201808,P10-13_adassxxviii}. The continued growth of the VO system requires that future and current data producing projects and missions are fully engaged with IVOA to ensure that the standards address the changing scientific needs of the community. The IVOA CSP, which has recently expanded with new members has the role of  fostering these engagements and identifying the most important scientific priorities to guide IVOA developments.  

Astronomy is rapidly entering into a new era of Big Data, and interoperability is increasingly important for multi-wavelength, multi-messenger and time domain astronomy. The participation of projects such as SKA, CTA, LSST, EUCLID etc., and the astronomy community in general, is essential to help define the scientific priorities that guide the IVOA activities in these areas. One of the common themes that is being driven by the needs of these big data projects is the concept of \emph{science analysis platforms} that will enable analysis of the data with capabilities for providing computational resources close to the data. Access to VO resources via these platforms will be essential, and the role of standardisation for interoperability of these platforms is a strongly emerging topic being discussed within IVOA. The user-centric focus of the science platforms presents many opportunities to enable a greater level of programmatic access to VO resources, with popular languages and systems used by scientists (e.g., Python notebooks). Other strong themes for future development include the use of machine learning, new visualisation capabilities, and use of VO data in education and outreach.

IVOA seeks to welcome new projects and their scientific and technical input. The IVOA web pages and wiki are being updated with information to facilitate the use of VO systems for data providers and astronomers, highlighting the many scientific and practical benefits to publishing data in the VO -- that the data becomes interoperable with other worldwide resources and that the data becomes visible in many widely used tools. For developers there is information on libraries, clients and tools that can be used to interact with the data when they are available through VO protocols. The revised web pages are also being updated with information on a number of well-tested frameworks available for publishing data.


Following this short-format interoperability meeting in College Park, the IVOA work continues via various email and slack communication channels, using the IVOA wiki for organisation of all the activities. The IVOA Newsletters \citep{ivoa_newsletter_201808} provide latest news and results, and a calendar of events. The recently appointed IVOA Media Group continues to enhance the IVOA social media presence, and is making plans for a revised IVOA information web portal.  A major 5-day Interoperability Meeting is planned for May 2019 in Paris.

\acknowledgements The IVOA would like to thank the local organisers of both the IVOA Interoperability Meeting, and the ADASS XVIII conference. 

\bibliography{H2}  

\end{document}